\documentclass[]{spie}  %>>> use for US letter paper
\usepackage{amsmath,amsfonts,amssymb}
\usepackage{graphicx}
\usepackage[colorlinks=true, allcolors=blue]{hyperref}
\usepackage{soul}

\usepackage[referable]{threeparttablex}
\usepackage{booktabs, longtable}
\newcounter{tablenote}[table]

\graphicspath{{figures/}}

\title{MagAO-X: Commissioning Results and Status of Ongoing Upgrades }

\author[a]{Jared R. Males}
\author[a]{Laird M. Close}
\author[a]{Sebastiaan Y. Haffert} % no Leiden affil?
\author[a,b]{Maggie Y. Kautz}
\author[a,b]{Jay Kueny}
\author[a]{Joseph D. Long}
\author[a]{Eden McEwen}
\author[f]{Noah Swimmer}
\author[g]{John I. Bailey, III}
\author[a,b]{Warren Foster}
\author[f]{Benjamin A. Mazin}
\author[a]{Logan Pearce}
\author[a,b]{Joshua Liberman}
\author[a,b]{Katie Twitchell}
\author[e]{Alycia J. Weinberger}
\author[a,b,c,d]{Olivier Guyon}
\author[a,b]{Alexander D. Hedglen}
\author[a,b]{Avalon McLeod}
\author[a]{Roz Roberts}
\author[a,b]{Kyle Van Gorkom}
\author[a]{Jialin Li}
\author[a]{Isabella Doty}
\author[a,b]{Victor Gasho}

\affil[a]{Steward Observatory, University of Arizona}
\affil[b]{James C. Wyant College of Optical Sciences, University of Arizona}
\affil[c]{Subaru Telescope, National Astronomical Observatory of Japan}
\affil[d]{Astrobiology Center, National Institutes of Natural Sciences, Japan}
\affil[e]{Earth and Planets Laboratory, Carnegie Institution for Science}
\affil[f]{Department of Physics, University of California, Santa Barbara}
\affil[g]{Caltech Optical Observatories, California Institute of Technology, Pasadena}

\authorinfo{Further author information: (Send correspondence to J.R.M.)\\J.R.M.: E-mail: jrmales@arizona.edu}
\begin{document} 
\maketitle

\begin{abstract}
MagAO-X is the coronagraphic extreme adaptive optics system for the 6.5 m Magellan Clay Telescope.  We report the results of commissioning the first phase of MagAO-X.  Components now available for routine
observations include: the $>2$ kHz high-order control loop consisting of a 97 actuator woofer deformable mirror (DM), a 2040 actuator tweeter DM, and a modulated pyramid wavefront sensor (WFS); classical Lyot
coronagraphs with integrated low-order (LO) WFS and control using a third 97-actuator non-common path correcting (NCPC) DM; broad band imaging in g, r, i, and z filters with two EMCCDs; simultaneous differential
imaging in H$\alpha$; and integral field spectroscopy with the VIS-X module.  Early science results include the discovery of an H$\alpha$ jet, images of accreting protoplanets at H$\alpha$, images of young extrasolar giant planets in the optical,
discovery of new white dwarf companions, resolved images of evolved stars,
and high-contrast images of circumstellar disks in scattered light in g-band (500 nm).  We have commenced an upgrade program, called “Phase II”, to enable high-contrast
observations at the smallest inner working angles possible.  These upgrades include a new 952 actuator NCPC DM to enable coronagraphic wavefront control; phase induced amplitude apodization
coronagraphs; new fast cameras for LOWFS and Lyot-LOWFS; and real-time computer upgrades.  We will report the status of these upgrades and results of first on-sky testing in March-May 2024.
\end{abstract}


\section{INTRODUCTION}
\label{sec:intro}  % \label{} allows reference to this section

MagAO-X\cite{2018SPIE10703E..09M,2020SPIE11448E..4LM,2022SPIE12185E..09M} is an
``extreme'' adaptive optics (ExAO) system optimized for high-contrast
science at visible-to-near-IR wavelengths.  It is installed on the Magellan Clay
6.5 m telescope at Las Campanas Observatory (LCO), in Chile. Construction and
commissioning of MagAO-X was funded by the NSF MRI program.    Commissioning of
MagAO-X is now complete, and it is routinely offered to all Magellan partners
for observations. Here we report on the outcome of this first phase of MagAO-X.

MagAO-X was conceived with the ultimate goal of reflected light characterization
of nearby temperate exoplanets.
This will require achieving exquisite wavefront control on-sky. This motivates a
program of continued upgrades and testing dubbed ``Phase II''.  We next provide
a status update on these upgrades, including the installation of a new 952
actuator non-common path deformable mirror (DM), a high-speed
low-order wavefront sensor (LOWFS) camera, and a real-time computer (RTC)
upgrade.

When not at the telescope, MagAO-X also serves as a valuable testbed for future
technology, including for AO and segment phasing technologies for the Giant
Magellan Telescope (GMT).  We will also provide a
brief overview of these activities and recent results.

\section{INSTRUMENT OVERVIEW}
\label{sec:overview}

Here we briefly present the design of MagAO-X and its key specifications.
Previous SPIE contributions provide more detail.
\cite{2018SPIE10703E..09M,2018SPIE10703E..5AV,2018SPIE10703E..55H,2018SPIE10703E..4ZL,2018SPIE10703E..4YC,
2018SPIE10703E..2QK,2018SPIE10703E..2NR,2018SPIE10703E..21S,2018SPIE10703E..1TM,2018SPIE10706E..5OK,2018SPIE10703E..1EG,2020SPIE11448E..4LM,2020SPIE11448E..0UC,
2022SPIE12185E..09M}
In addition, the complete preliminary design review (PDR) documentation is available at
\url{https://magao-x.org/docs/handbook/appendices/pdr/}, and results of laboratory integration and preparation for shipment can be bound in the pre-ship review (PSR)
documentation:  \url{https://magao-x.org/docs/handbook/appendices/psr/index.html}.

\begin{figure}[h]
   \centering
   \includegraphics[width=2.5in]{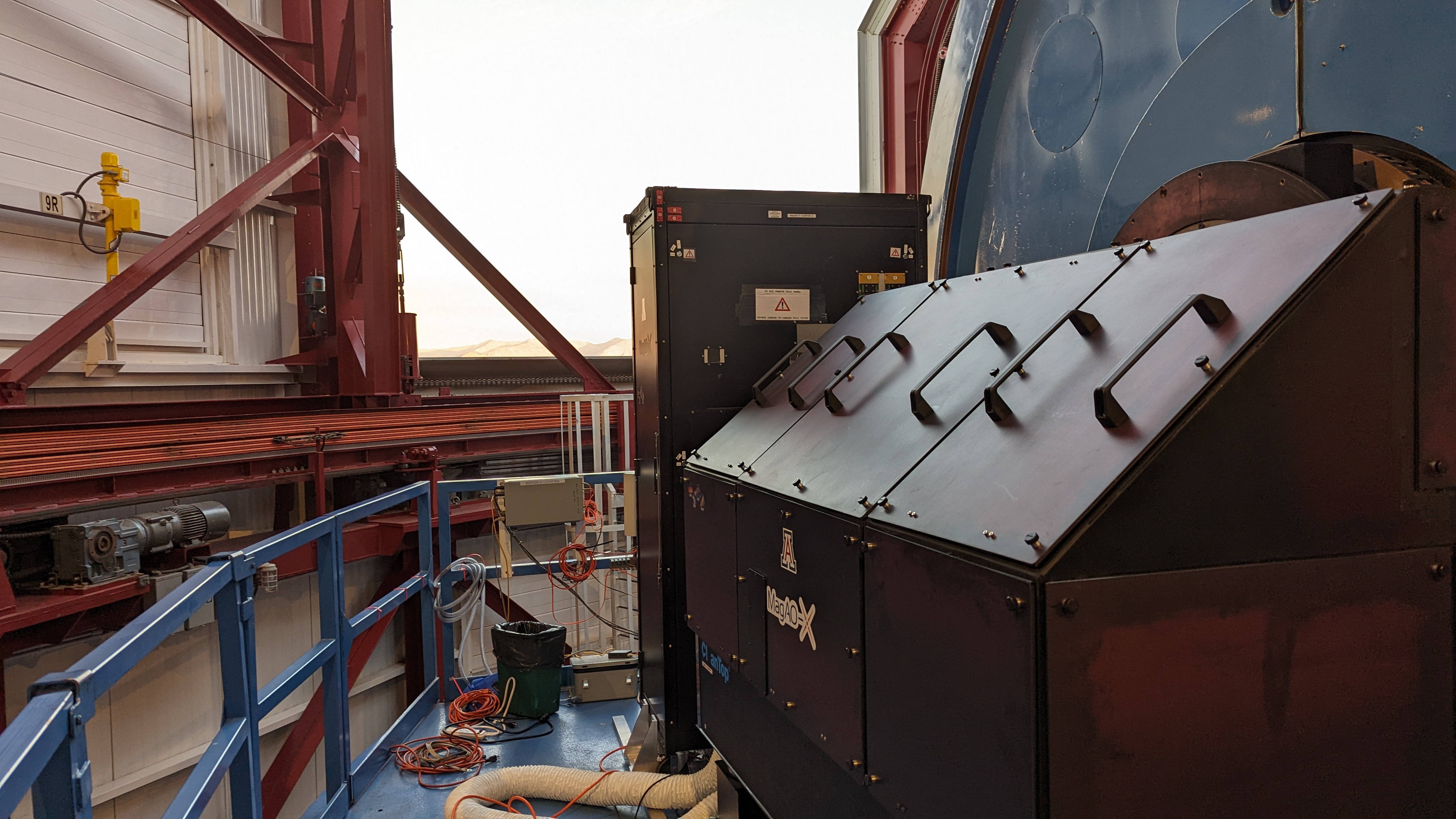}
   \includegraphics[width=4.0in]{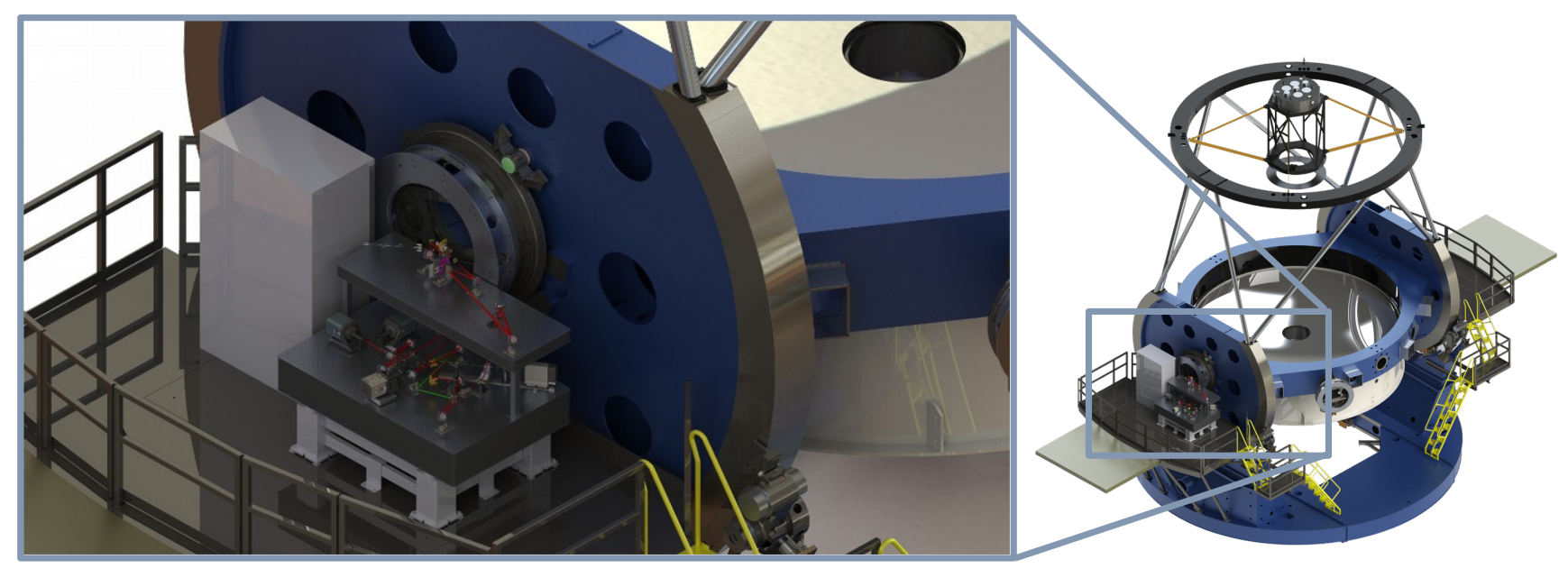}
   \caption{When observing at LCO, MagAO-X occupies the Nasmyth platform of the Magellan Clay 6.5 m telescope. \label{fig:platform_inset}}
\end{figure}

\begin{figure}[t]
   \centering
   \includegraphics[width=3.25in]{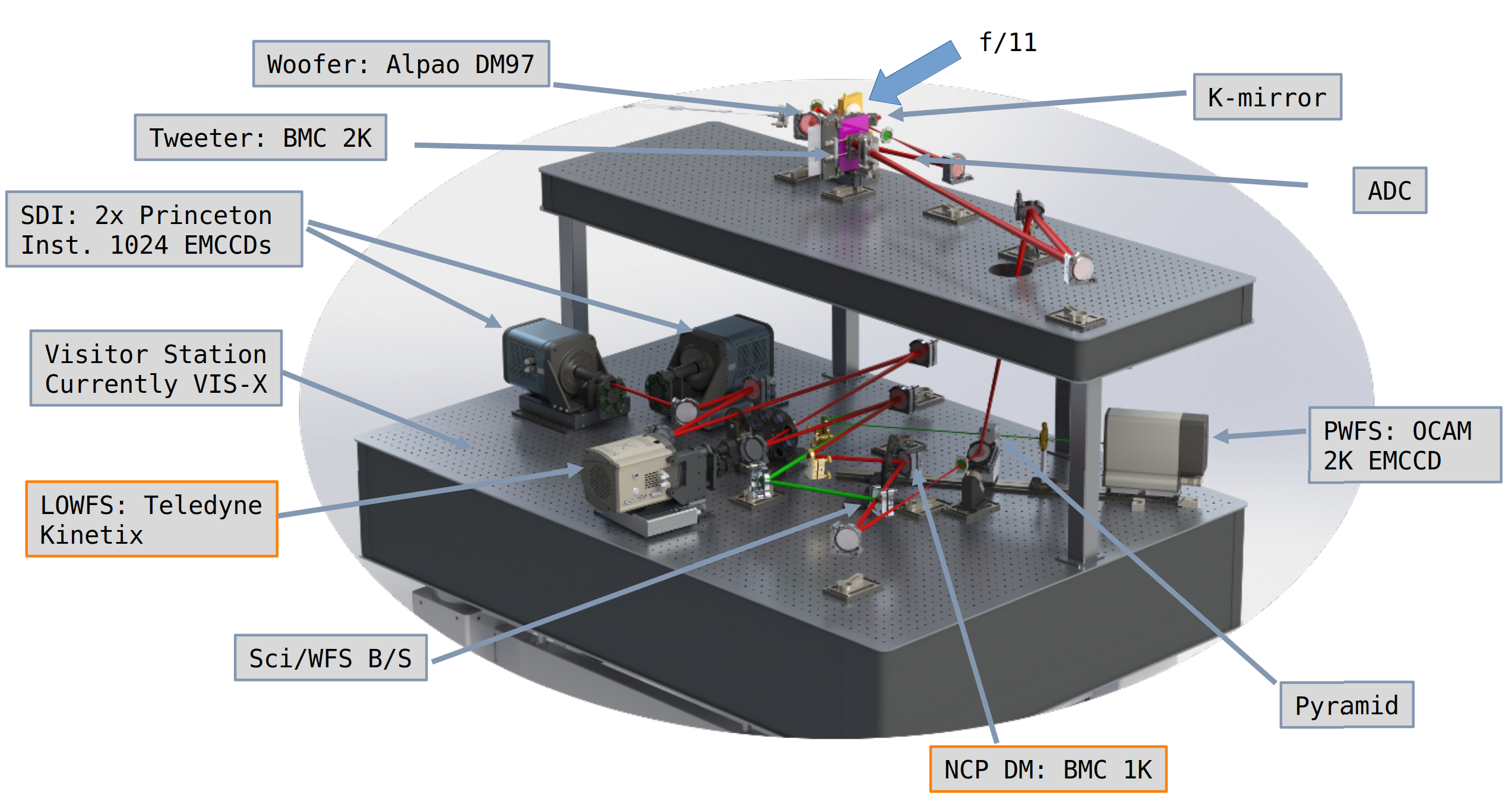}
   \includegraphics[width=3.25in]{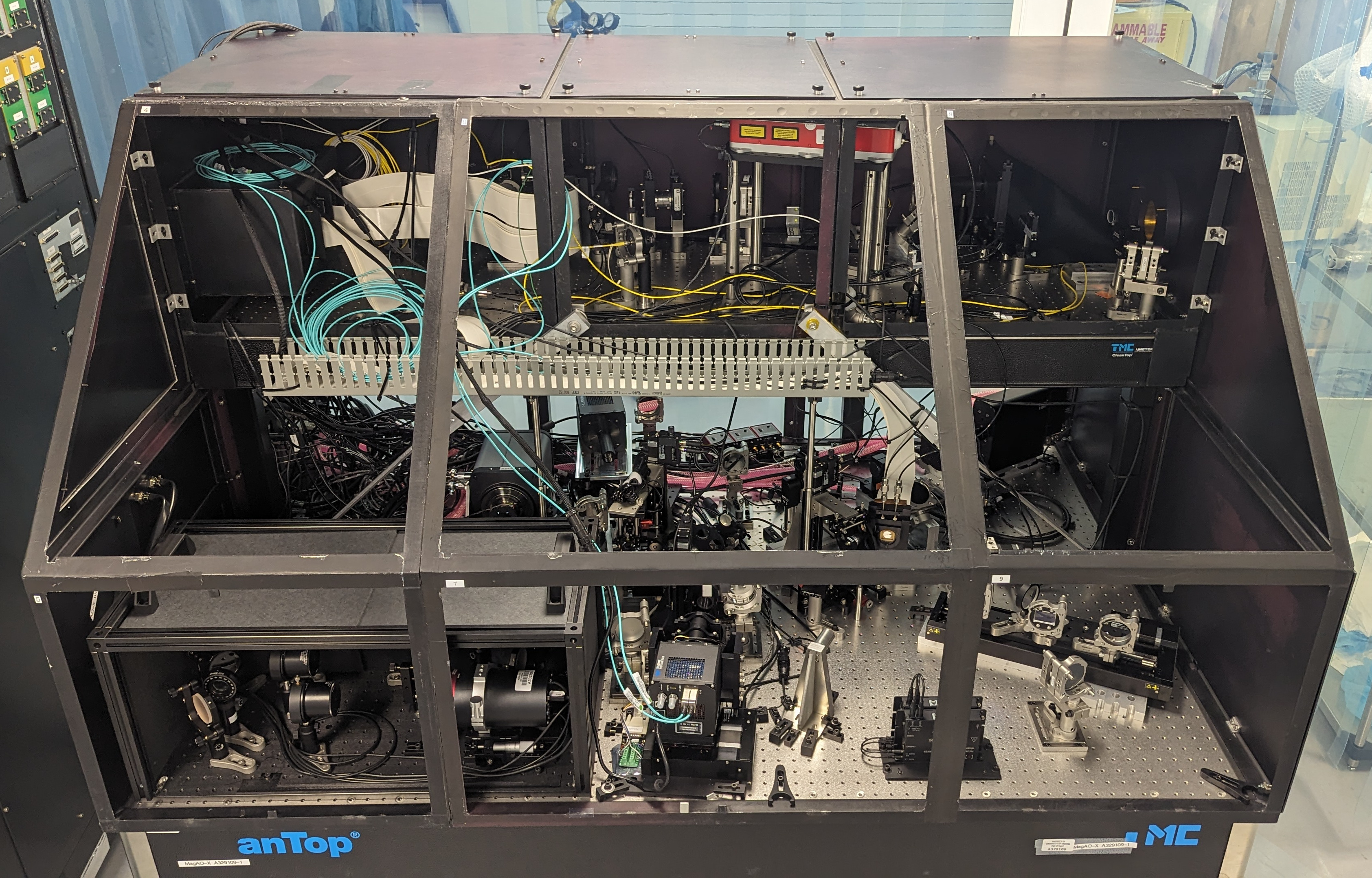}
   \caption{ Left: Overview of the MagAO-X opto-mechanical layout. Recently installed Phase II upgrades highlighted in orange.
             Right: as-built MagAO-X just prior to the May 2024 observing run. \label{fig:labeled}}
\end{figure}

MagAO-X is installed on the Nasmyth platform of the 6.5 Magellan Clay telescope (Figure \ref{fig:platform_inset}) at Las Campanas Observatory, Chile.
MagAO-X is a woofer-tweeter system, with a 97-actuator deformable mirror (DM) serving as the woofer and a 2040 actuator MEMS DM as the tweeter.
The pupil illuminates 48 actuators across in the short direction, giving a 13.5 cm actuator pitch, with approximately 1640 actuators illuminated.

MagAO-X is a two-level optical bench (Figure \ref{fig:labeled}).
The upper-level houses the k-mirror derotator, woofer, tweeter, and atmospheric dispersion compensator.  A periscope relay folds the beam to the lower level
containing the pyramid wavefront sensor (PyWFS).  A selectable beamsplitter sends light to the PyWFS and to a Lyot-coronagraph system feeding a dual-EMCCD
simultaneous differential imaging (SDI) system or optionally visitor instruments such as the VIS-X spectrograph\cite{2021SPIE11823E..06H}.
The coronagraph contains a third deformable mirror, originally with 97 actuators but now upgraded to 952, after the beamsplitter.  This
enables non-common path offloading without offsetting the PyWFS.  All of these components and subsystems are described in great detail in
the references and documents noted above.

Science cases motivating MagAO-X include:

\begin{enumerate}
\item The search for and characterization of young accreting planets in H$\alpha$ orbiting nearby T Tauri
and Herbig Ae/Be stars \cite{2020AJ....160..221C}.   See Li et al\cite{li_2024} in these proceedings for an update on this project.

\item Circumstellar disk characterization

\item Young EGP characterization in the red-optical/near-IR

\item Spatially resolved stellar surface and evolved star imaging at high spectral resolution.

\item Characterization of tight binary star systems.

\item \textit{Kepler} and \textit{TESS} followup: MagAO-X can be used for high spatial-resolution follow-up of potential transiting planet host
stars to search for binarity and other sources of contamination.

\item Accelerating Stars: Stars exhibiting accelerations measured by astrometry and radial velocity are proving to be excellent places to
search for low-mass companions. Led by S. Haffert and L. Pearce, the MagAO-X Xoomies survey is conducting a cued search of such stars.

\item White Dwarf Companions: MagAO-X is sensitive to an understudied population of white dwarf companions to nearby main sequence stars (so-called Sirius-like systems).  The ExAO Pup Search
is characterizing such systems (Pearce et al, in prep).

\item High spatial resolution imaging of asteroid surfaces and asteroid companion searches.

\end{enumerate}

\begin{table}[t]
\begin{ThreePartTable}
\begin{TableNotes}
\item[a] \label{tn:a} $\sim$0.5 night used each time for post-installation checkout, here included in commissioning and Ph. II allocations
\item[b] \label{tn:b} Nights allocated for commissioning of VIS-X
\item[c] \label{tn:c} Nights allocated for commissioning Phase II upgrades
\item[d] \label{tn:d} Denotes whether MagAO-X was returned to Tucson, AZ, following the run
\item[e] \label{tn:e} 2024A was scheduled in two distinct runs
\end{TableNotes}
\begin{longtable}{c|c|c|c|c|c|c|l}
\caption{On-sky nights for MagAO-X. \label{tab:nights}} \\
                 &                  & \multicolumn{4}{c|}{Nights}                             & Return            & \\
Run              & Dates            & Tot. & Com.\tnote{a} & VIS-X\tnote{b} & Ph. II\tnote{c} & Shipment\tnote{d} &  Notes \\
\hline
\hline
2019B            & Dec 2 -- Dec 8   & 4    & 4             &                &                 & Y                 & two separate 2-night blocks \\
2022A            & Apr 9 -- Apr 23  & 15   & 5             &  1             &                 & Y                 & \\
2022B            & Nov 28 -- Dec 9  & 12   & 2             &  1             &                 & N                 & significant weather losses\\
2023A            & Mar 3 -- Mar 16  & 14   & 2             &                &  1              & Y                 & XKID commissioning \\
2024Aa\tnote{e}  & Mar 18 -- Apr 1  & 13.5 &               &                &  2              & N                 & 1 night shared\\
2024Ab\tnote{e}  & May 14 -- May 24 & 11   &               &                &  1              & N                 & $>80$\% lost due to weather\\
\hline
\multicolumn{2}{r}{total:}          &  69.5 & 13            &  2             &  4              &                   &\\
\insertTableNotes \\
\end{longtable}
\end{ThreePartTable}
\end{table}

\section{COMMISSIONING RESULTS}
\label{sec:commissioning}

Commissioning of the initial MRI-funded capabilities of MagAO-X is now complete. Table \ref{tab:nights} summarizes the on-sky time utilized by MagAO-X on the
Clay Telescope.  Starting with the 2nd commissioning run (2022A) we have made
significant time available for partner observations.  13 nights out of 69.5 were
used for explicit commissioning tasks.  Additionally 2 nights have been used
for VIS-X IFU commissioning and 4 nights so far have been devoted to
commissioning Phase II upgrades.

\clearpage

\subsection{XKID}

\begin{figure}[t]
   \centering
   \includegraphics[width=2.5in]{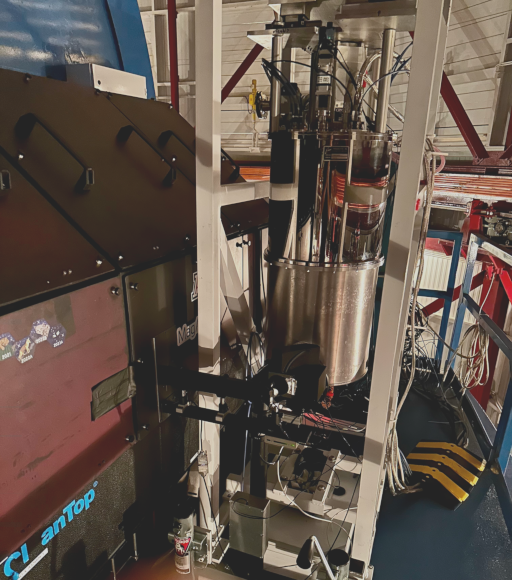}
   \caption{The XKID IFU dewar aligned to MagAO-X. \label{fig:xkid}}
\end{figure}

In March, 2023, we commissioned the MagAO-X Kinetic Inductance detector (XKID) IFU behind MagAO-X.
XKID is a superconducting integral field spectrograph designed for
low spectral resolution, photon counting spectroscopy of exoplanets in the optical and near-IR
(800-1400 nm) based on an MKIDs array\cite{2020PASP..132l5005W}. The high time resolution (microsecond photon counting) and lack of detector
noise (zero read noise or dark current) allows significant reduction of speckle backgrounds
through the Stochastic Speckle Discrimination (SSD) technique \cite{2019PASP..131k4506W,2021AJ....162...44S,2022AJ....164..186S}.  Figure \ref{fig:xkid}
shows the XKID dewar aligned to MagAO-X on the Clay platform.

\subsection{VIS-X}

VIS-X is a visible wavelength IFU spectrograph in MagAO-X, see Haffert et al.\cite{2021SPIE11823E..06H}.
VIS-X supports three modes: the wide-field ``Emission Line Mode'' ($\mathcal{R}=$15,000, 653-658 nm, 1.15''X1.15'') and the wide-field ``Low-res Mode''
($\mathcal{R}=$50--100, 450-900 nm, 1.15''X1.15''), and narrow-field ``Characterization Mode'' ($\mathcal{R}=$15,000, 500-900nm, 0.063''X0.5'').
The ``Emission Line Mode`` has been commissioned and has been used to observe young stellar objects with jets, accreting companions,
resolved stellar surfaces and (interacting) close binaries in the H$\alpha$ emission line (Haffert et al, in prep).
Final commissioning of the remaining two modes is planned for the Fall 2024 semester.

\section{PHASE II UPGRADES}
\label{sec:upgrades}

\subsection{NCPC DM}
The correction of non-common path aberrations (NCPA) is critical to achieving the highest image quality in the science focal plane of an AO system.
In typical AO systems the only corrective elements are prior to the beamsplitter, and so are in common-path with the WFS and the
science instrumentation.  The challenge this creates is that any aberrations non-common between the WFS and science instrument must
be corrected using the upstream DM(s), which then requires that offsets be applied to the WFS signal such that the control loop does
not simply remove the desired correction.

The solution we arrived at in MagAO-X is to place a non-common path correcting DM, which we refer to as ``dmncpc'', after the beamsplitter.  This DM
can then be used to correct NCPA without disturbing the upstream high-order loop.  The original version of MagAO-X included a 97 actuator device
identical to the woofer.  In Phase II we have upgraded this to a 952 actuator MEMS device.  We make use of focus diversity phase retrieval\cite{2018SPIE10703E..5AV}
(FDPR), as well as coronagraphic low-order wavefront sensing using defocused spots to sense and correct NCPA.  See Kueny et al.\cite{kueny_2024} (these proceedings) for
details about the characterization and on-sky commissioning of the new 1K dmncpc.  See Haffert et al.\cite{haffert_2024} (these proceedings) for details on how this
upgrade has revolutionized our post-coronagraph high-contrast imaging capabilities, and Liberman et al.\cite{liberman_2024} (these proeceedings) for an analysis of
the NCPC DM and coronagraph alignment tolerances.

\begin{figure}[h]
   \centering
   \includegraphics[width=3.25in]{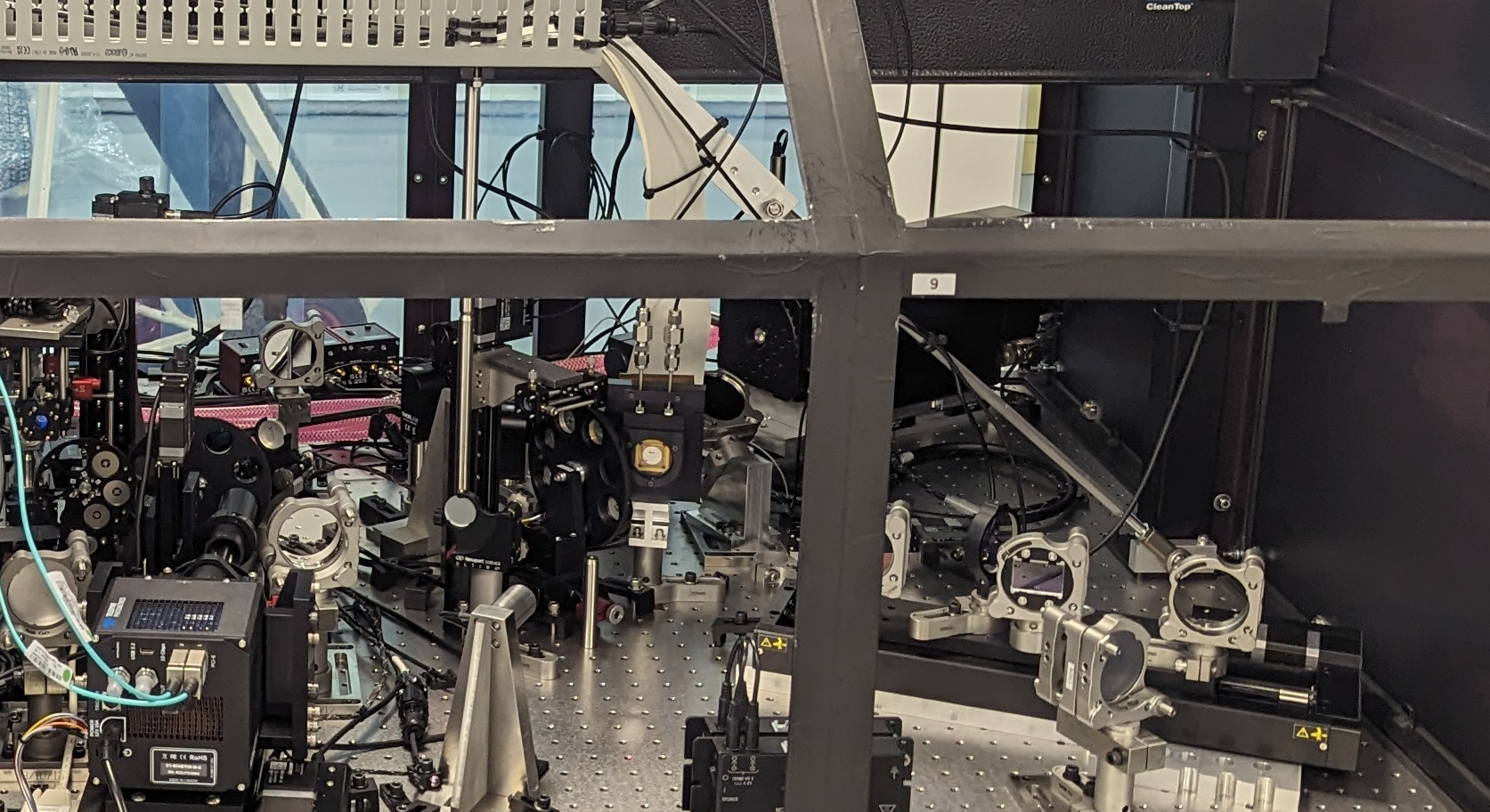}
   \caption{New NCPC DM \label{fig:dmncpc}}
\end{figure}

\subsection{RTC}

The control system for MagAO-X is based on commercial-off-the-shelf (COTS) computing components, including standard ATX motherboards, CPUs, and consumer-grade GPUs.
This enables us to upgrade the performance of MagAO-X as computer hardware improves.  The original RTC of MagAO-X was based on dual Intel Xeon 18 core processors on a
PCIe Gen 3 motherboard, and
utilized three NVIDIA 2080 Ti GPUs on an PCIe expansion board.  As part of Phase II we upgraded to the AMD Threadripper architecture, with a 64 core processor on a
PCIe Gen 4 motherboard. Additionally two GTX 4090 GPUs replaced the 2080s.  This improved our total reconstruction time from $\sim$200 $\mu$sec to $\sim$75 $\mu$sec.
The move to PCIe Gen 4 also resulted in significantly improved timing stability, as we now have many fewer PCIe bus collisions between the WFS images, DM commands,
and GPU memory operations.

\begin{figure}[h]
   \centering
   \includegraphics[width=1.75in]{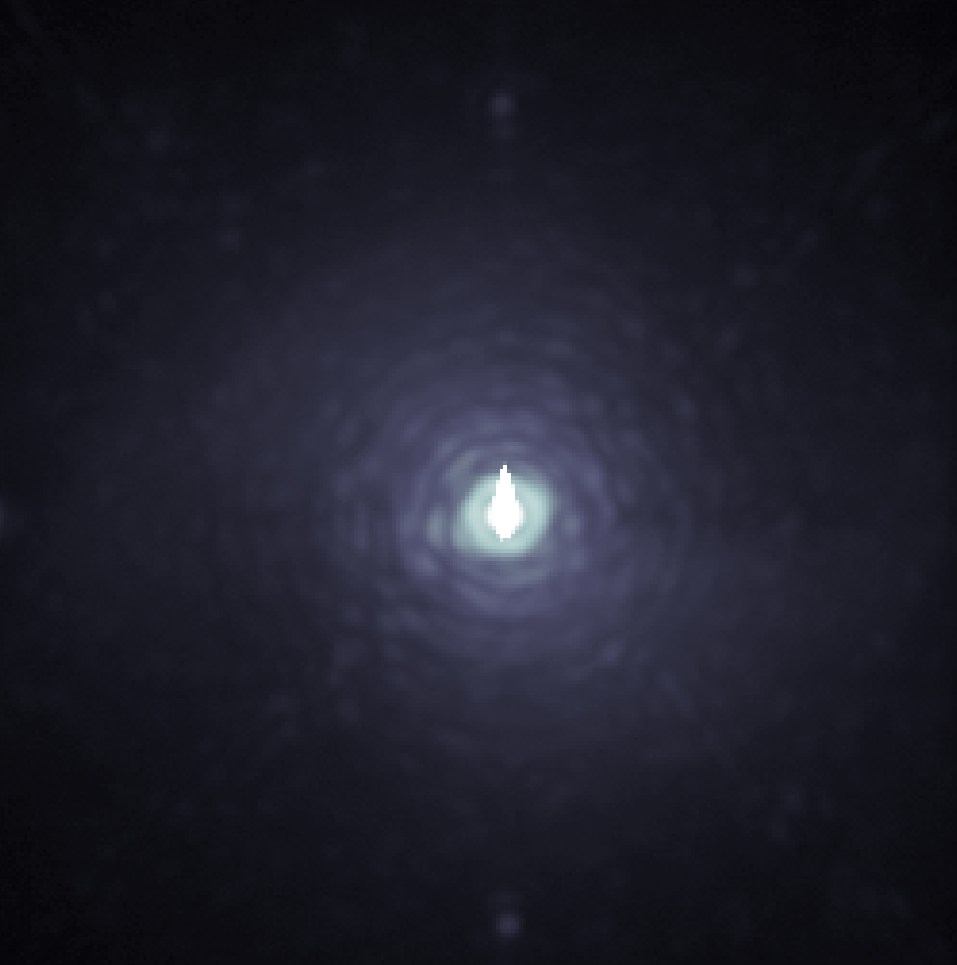}
   \includegraphics[width=1.75in]{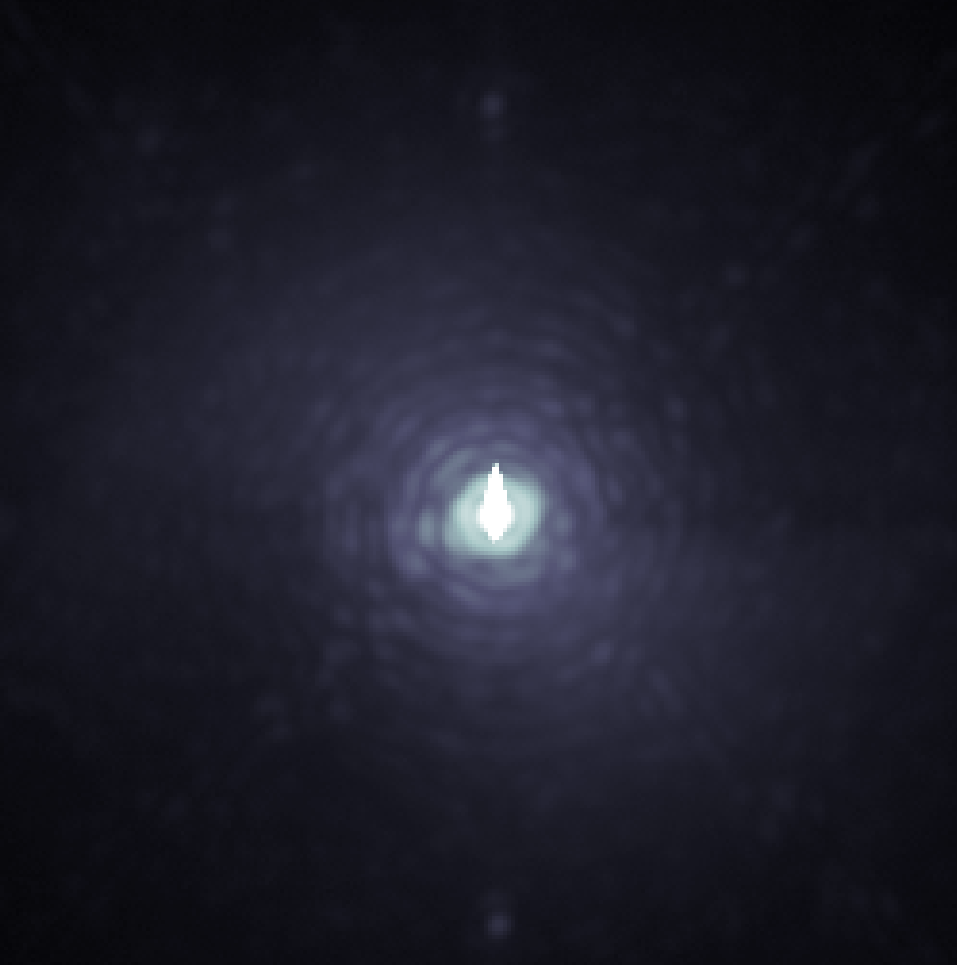}
   \includegraphics[width=3in]{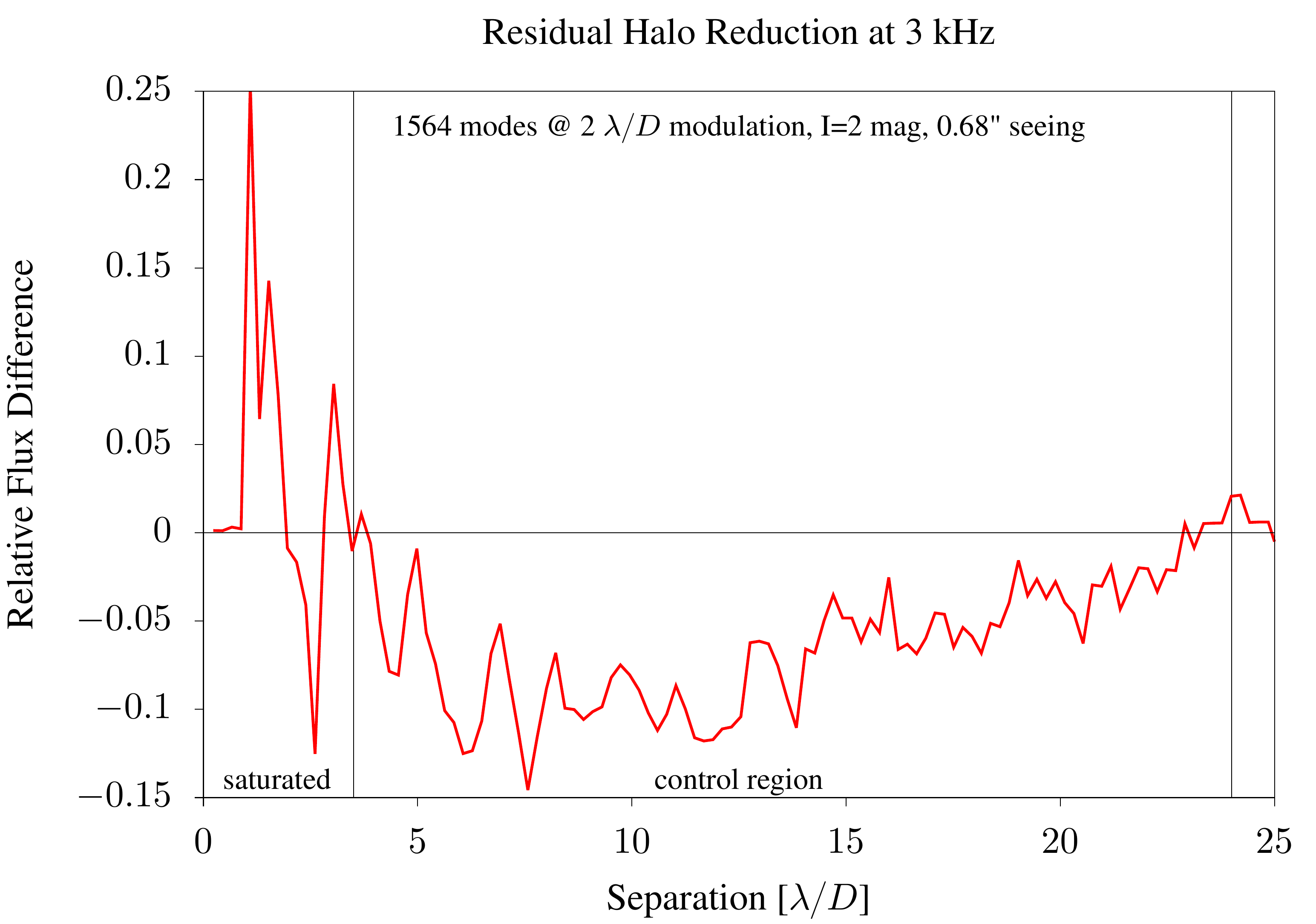}
   \caption{The RTC upgrade enabled achieving a 3 kHz loop speed.  Shown at left is the PSF at 2 kHz, which compares to the middle PSF at 3 kHz.  The middle image is
   more saturated in the core and improvements in the residual wind-driven halo are evident.  The right-hand plot shows the improvement in residual flux at 3 kHz
   compared to 2 kHz. \label{fig:3khz}}
\end{figure}

The computer hardware upgrade enabled us to achieve 3 kHz loop speeds on-sky for the first time.  The resulted in an over 10\% reduction in residual flux,
see Figure \ref{fig:3khz}.

\subsection{PIAA}
\begin{figure}[t]
   \centering
   \includegraphics[width=2in]{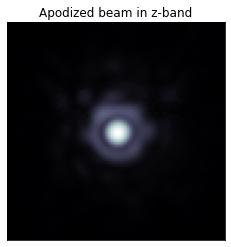}
   \includegraphics[width=2in]{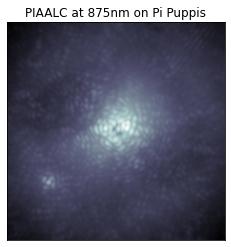}
   \caption{On-sky results with Phase Induced Apodization (PIAA) lenses.  Left image shows the on-axis PSF with no focal plane mask (FPM), demonstrating that the PIAA lenses are aligned.
            Right image shows the result with a opaque Lyot Coronagraph (LC) FPM in the beam.  The off-axis companion has only a small amount of residual astigmatism, showing that
            the inverse PIAA lenses are also well aligned. \label{fig:piaa}}
\end{figure}

A further upgrade underway is the implementation of Phase Induced Amplitude Apodization Complex Mask Coronagraphs (PIAACMC) in MagAO-X.  We have designed, procured, and
installed a set of PIAA and inverse PIAA lenses and conducted initial tests on-sky\cite{FosterThesis2023}.  Results with an opaque focal plane mask (FPM) are shown in Figure \ref {fig:piaa}.
We are finalizing designs for transmissive complex FPMs to complete the PIAACMC implementation for demonstration on-sky in Fall, 2024.

\subsection{LOWFS Upgrades}

To take advantage of the high-speed capabilities of the new NCPC DM and to prepare for using transmissive complex masks in PIAACMC, we have upgraded our low-order WFS (LOWFS) camera
to use the Teledyne Kinetix camera.  Over the small FOVs used for LOWFS on MagAO-X\cite{McLeodThesis2023} this device is capable of over 10,000 frames per second.   In addition to
upgrading the existing FPM based LOWFS, we have also begun implementing a second Lyot-based LOWFS\cite{2015PASP..127..857S} system using a second Kinetix camera.
We plan to demonstrate this capability in Fall, 2024.

\subsection{Algorithm Development}

We continue to optimize the control system.  Ongoing efforts include machine-learning driving non-linear PyWFS reconstruction algorithms (Landman et al\cite{landman_2024}, these proceedings), and PyWFS ``optical gain'' characterization and on-line correction (McEwen et al.\cite{mcewen_2024}, these proceedings).  Efforts are also underway to implement real-time atmospheric dispersion corrector (ADC) control (Twitchell et al.\cite{twitchell_2024}, these proceedings). Finally, we are working to implement PSF/speckle reconstruction using system telemetry, see Long et al.\cite{long_2024} (these proceedings) for recent results.

\section{TESTBED OPERATIONS}

As part of the plan to continuously improve MagAO-X, it was designed for routine shipment between LCO and the University of Arizona in Tucson.  This has enabled MagAO-X to serve as a key part of the High-Contrast
Adaptive Optics Testbed (HCAT) for the Giant Magellan Telescope (GMT), part of the overall program to develop segment phase sensing and control strategies for the GMT\cite{Demers_2022}.  MagAO-X serves as the
simulated GMT-AO system for HCAT, and is used to conduct phase sensing and control experiments.\cite{2020SPIE11448E..2XH,Hedglen_2022,Kautz_2022,10.1117/1.JATIS.8.2.021513,10.1117/1.JATIS.8.2.021515}.  For more
details on these experiments see Close et al.\cite{close_2024}, Quiros-Pacheco et al.\cite{pacheco_2024}, Plantet et al.\cite{plantet_2024}, and Demers et al.\cite{demers_2024} in these proceedings, and Kautz et al. (in prep).

As of July, 2024 MagAO-X is set up at LCO for remote operations.  It is available for WFS\&C experiments and algorithm development.  We expect it to remain at LCO through April, 2025.

\section{CONCLUSION}
\label{sec:conclusion}

MagAO-X has now completed initial commissioning at LCO, and is supporting science observations by Magellan partner scientists.  Significant improvements in performance have been realized over the commissioning period which began in Dec, 2019.  In addition, ongoing upgrades to MagAO-X and its ability to serve as a WFS\&C testbed will continue to improve performance.

\acknowledgments % equivalent to \section*{ACKNOWLEDGMENTS}       
 
We are very grateful for support from the  NSF MRI Award \#1625441 (MagAO-X).
The Phase II upgrade program is made possible by the generous support of the
Heising-Simons Foundation.

% References
\bibliography{report} % bibliography data in report.bib
\bibliographystyle{spiebib} % makes bibtex use spiebib.bst

\end{document}